\documentclass[aps, prd, amsmath, floats,floatfix, twocolumn, superscriptaddress]{revtex4}

\usepackage[normalem]{ulem}
\usepackage{amssymb}
\usepackage{amsmath}
\usepackage{verbatim}
\usepackage{mathrsfs}
\usepackage{amsfonts}
\usepackage{latexsym}
\usepackage{epsfig}
\usepackage{color}
\usepackage{graphicx,subfigure}
\usepackage{grffile}
\usepackage{units}
\usepackage{slashbox}
\usepackage{envmath}
\usepackage{natbib}
\usepackage{hyperref}
\usepackage[utf8]{inputenc}

\begin{document}


\definecolor{orange}{rgb}{0.9,0.45,0}

\newcommand{\re}{\mbox{Re}}
\newcommand{\im}{\mbox{Im}}

\newcommand{\tf}[1]{\textcolor{red}{#1}}
\newcommand{\nsg}[1]{\textcolor{blue}{NSG: #1}}
\newcommand{\ch}[1]{\textcolor{green}{CH: #1}}
\newcommand{\fdg}[1]{\textcolor{orange}{FDG: #1}}
\newcommand{\pcd}[1]{\textcolor{magenta}{#1}}
\newcommand{\mz}[1]{\textcolor{cyan}{[\bf MZ: #1]}}

\def\CovDev{D}
\def\Res{{\mathcal R}}
\def\Gammaflat{\hat \Gamma}
\def\metricflat{\hat \gamma}
\def\Dflat{\hat {\mathcal D}}
\def\part_n{\partial_\perp}

\def\Lie{\mathcal{L}}
\def\A{\mathcal{X}}
\def\Aphi{\A_{\phi}}
\def\hAphi{\hat{\A}_{\phi}}
\def\E{\mathcal{E}}
\def\Ham{\mathcal{H}}
\def\M{\mathcal{M}}
\def\R{\mathcal{R}}
\def\p{\partial}

\def\hg{\hat{\gamma}}
\def\hA{\hat{A}}
\def\hD{\hat{D}}
\def\hE{\hat{E}}
\def\hR{\hat{R}}
\def\hcA{\hat{\mathcal{A}}}
\def\hDelt{\hat{\triangle}}

\def\be{\begin{equation}}
\def\ee{\end{equation}}

\renewcommand{\t}{\times}

\long\def\symbolfootnote[#1]#2{\begingroup%
\def\thefootnote{\fnsymbol{footnote}}\footnote[#1]{#2}\endgroup}


\title{Can fermion-boson stars reconcile multi-messenger observations of compact stars?}

  \author{Fabrizio Di Giovanni}
\affiliation{Departamento de
  Astronom\'{\i}a y Astrof\'{\i}sica, Universitat de Val\`encia,
  Dr. Moliner 50, 46100, Burjassot (Val\`encia), Spain}

\author{Nicolas Sanchis-Gual}
\affiliation{Departamento  de  Matem\'{a}tica  da  Universidade  de  Aveiro  and  Centre  for  Research  and  Development in  Mathematics  and  Applications  (CIDMA),  Campus  de  Santiago,  3810-183  Aveiro,  Portugal}

 \author{Pablo Cerd\'a-Dur\'an
}
\affiliation{Departamento de
  Astronom\'{\i}a y Astrof\'{\i}sica, Universitat de Val\`encia,
  Dr. Moliner 50, 46100, Burjassot (Val\`encia), Spain}  

\author{Jos\'e~A. Font}
\affiliation{Departamento de
  Astronom\'{\i}a y Astrof\'{\i}sica, Universitat de Val\`encia,
  Dr. Moliner 50, 46100, Burjassot (Val\`encia), Spain}
\affiliation{Observatori Astron\`omic, Universitat de Val\`encia,  Catedr\'atico Jos\'e Beltr\'an 2, 46980, Paterna (Val\`encia), Spain}


\date{\today}


\begin{abstract} 
Mixed fermion-boson stars are stable, horizonless, everywhere regular solutions of the coupled Einstein-(complex, massive) Klein-Gordon-Euler system. While isolated neutron stars and boson stars are uniquely determined by their central energy density, mixed configurations conform an extended parameter space that depends on the combination of the number of fermions and (ultra-light) bosons. The wider possibilities offered by fermion-boson stars could help explain the tension in the measurements of neutron star masses and radii reported in recent multi-messenger observations and nuclear-physics experiments. In this work we construct  equilibrium configurations of mixed fermion-boson stars with realistic equations of state for the fermionic component and different percentages of bosonic matter. We show that our solutions are in excellent agreement  with  multi-messenger data, including gravitational-wave events GW170817 and GW190814 and  X-ray  pulsars  PSR  J0030+0451  and PSR J0740+6620, as well as with nuclear physics constraints from the PREX-2 experiment.
\end{abstract}

\maketitle

\vspace{0.8cm}

\bigskip

\section{Introduction}
 The determination of the equation of state (EoS) of matter at the supernuclear densities attained in neutron star interiors is a long-standing issue in nuclear astrophysics (see~\cite{Lattimer:2012,Lattimer:2016} and references therein). High-precision measurements of the masses and radii of neutron stars are necessary to confidently constrain the EoS. Recent observations in both the electromagnetic channel and the gravitational-wave channel, together with constraints from nuclear physics, are helping shed light on this issue, yet uncertainties remain~\cite{GW170817,Margalit:2017,Riley:2019,Miller:2019,Dietrich:2020,Landry:2020,Katerina:2020,Riley:2021,Miller:2021,Raaijmakers:2021,Breschi:2021}. 
 
 During the last decade it has been possible to accurately measure the mass of two milisecond pulsars with masses close to $2~M_\odot$,  PSR J1614-2230 \cite{Demorest:2010,Fonseca:2021} and PSR J0348+0432 \cite{Antoniadis:2013}. 
 These results impose a strong lower limit to the maximum mass of neutron stars and have constrained considerably the properties of dense matter \cite{Lattimer:2016}. However, only recently it has been possible an accurate joint determination of the mass {\it and} the radius of a neutron star.  Bayesian inference on pulse-profile modeling of observations from the Neutron Star Interior Composition Explorer (NICER) of the rotation-powered, X-ray milisecond pulsar PSR J0030+0451, yielded values of the mass and (circumferential) radius of  $\sim 1.4\,M_{\odot}$ and $\sim 13\,$km, respectively~\cite{Miller:2019,Riley:2019}. Even more recently, the same teams of researchers have reported the joint determination of the mass and radius of PSR J0740+6620~\cite{Riley:2021,Miller:2021}, the most massive known neutron star. Combining data from NICER and XMM-Newton~\cite{Miller:2021}, and also accounting for radio timing (Shapiro delay) in the case of~\cite{Riley:2021} (see also~\cite{Fonseca:2021}), these teams have inferred values for the mass and radius of $2.08~M_\odot$ and $\sim 13$~km, respectively. The fact that J0740+6620 is about 50\% more massive than J0030+0451 while both objects are  essentially the same size challenges theoretical models of neutron-star interiors. 

Gravitational waves have also been able to put joint constraints on the neutron star mass and radius. The first-ever detection of a binary neutron star merger by the LIGO-Virgo Collaboration (LVC), GW170817 \cite{TheLIGOScientific:2017qsa}, made it possible not only to place constraints on the individual masses of the components of the binary but also on the tidal deformability of neutron stars, which has been used to constrain the neutron star radius~\cite{GW170817} (see also~\cite{Katerina:2020,Dietrich:2021} and references therein).

In addition, the interpretation of the recent LVC detection of the compact binary merger event GW190814~\cite{GW190814:2020} poses quite some difficulties. While the mass of the primary component, $23.2\,M_{\odot}$, allows to conclusively identify it as a black hole, the mass of the secondary, $2.50-2.67\,M_{\odot}$, raises doubts on the nature of this component, which might be either a black hole or a neutron star. If the latter were the case it would be the most massive neutron star ever observed. A number of recent investigations have tried to explain such a large mass~\cite{Most:2020,Nathanail2021,Sedrakian:2020,Tsokaros:2020,Biswas:2021,Zhang:2020,Bombaci:2021,Dexheimer:2021,Godzieba:2021,Vattis:2020,Cholis:2021,Charmousis:2021,Das:2021,Lee:2021}. Proposals include the possibility that the secondary were a rapidly-rotating neutron star that collapsed to a spinning black hole before merger~\cite{Most:2020,Nathanail2021}, a neutron star with a stiff high-density EoS or a sufficiently large spin~\cite{Biswas:2021} (see also~\cite{Zhang:2020}), or a neutron star with exotic degrees of freedom, i.e.~a strange quark star, within the scenario in which neutron stars and quark stars coexist~\cite{Bombaci:2021} (see also~\cite{Dexheimer:2021}). Somewhat more exotic possibilities involving slowly-rotating neutron stars in 4D Einstein-Gauss-Bonnet gravity~\cite{Charmousis:2021}, primordial black holes~\cite{Vattis:2020}, Thorne-$\dot{\rm Z}$ytkow objects~\cite{Cholis:2021}, or dark-matter-admixed neutron stars~\cite{Das:2021,Lee:2021}, have also been suggested.

The neutron star radius can also be constrained by improving the measurement of nuclear interaction parameters \cite{Lattimer:2012,Drischler:2020}.
Very recently the PREX-2 experiment has measured with high accuracy the neutron skin thickness of $^{208}$Pb~\cite{PREX2} which constrains the neutron star radius for a $1.4~M_\odot$ neutron star to be larger than $13.25$~km \cite{Reed:2021}. Although compatible with milisecond pulsar radius measurements, this result is in some tension with the gravitational-wave determinations~\cite{Essick:2021}. The combined constraints of the multi-messenger data and PREX-2 measurements have been shown by~\cite{Li:2021,Tang:2021} to be compatible with models of hybrid stars with first-order phase transition from nucleonic to quark matter in the core, a result disfavored by the analysis of~\cite{Pang:2021}. 

Additionally, the nuclear physics modeling of realistic EoS at high densities has lead to the so-called hyperon problem [see e.g.~\cite{Bedaque:2015} and references therein]. In order to reach the high masses necessary to fulfill the observational constraints on the maximum mass of neutron stars, models have to reach high central densities, at which the appearance of hyperons is expected. However, the presence of hyperons may soften the EoS at those densities and limit the possible values for the maximum mass, making it difficult to reach the $\sim2~M_\odot$ constraint.

Motivated by these observational and experimental results we put forward in this paper a theoretically-motivated new model based on mixed fermion-boson stars, i.e.~neutron stars that incorporate some amount of bosonic matter. Using this model we are able to construct existence plots (mass-radius equilibrium configurations) compatible with multi-messenger observational data, including gravitational-wave events GW170817 and GW190814, and X-ray pulsars PSR J0030+0451 and PSR J0740+6620. We note that our model shares some similarities with those of~\cite{Das:2021,Lee:2021} but also some differences. The study of \cite{Das:2021} is only focused on GW190814 and explains the mass of the secondary by admixing neutron stars modelled by stiff EoS with nonannihilating weakly interacting massive particles dark matter. On the other hand, the very recent study of~\cite{Lee:2021} also focuses only on GW190814 and explains the mass of the secondary by resorting to a neutron star admixed with at least $2.0\,M_\odot$ dark matter made of axion-like particles. In our study (see below) we employ a complex scalar field while in~\cite{Lee:2021} the authors consider a real field to model QCD axions. 

Ultralight bosons form localized, coherently oscillating configurations
akin to Bose-Einstein condensates~\cite{Sin:1992bg, Chavanis:2011cz}. For light-enough bosonic particles, i.e.~with a mass $\mu\sim 10^{-22}\,$eV, these condensates have been proposed to explain large-scale structure formation through dark-matter seeds~\cite{Matos:1999et,Hu:2000ke}. Heavier bosons lead to much smaller configurations with the typical size and mass of neutron stars --  hence the name boson stars~\cite{Kaup:1968zz,Ruffini:1969qy} (see~\cite{jetzer:1992,liebling2017dynamical} and references therein). It is worth mentioning that 
recent examples have shown the intrinsic degeneracy between the prevailing Kerr black hole solutions of general relativity and boson-star solutions, using both gravitational-wave data~\cite{CalderonBustillo:2020srq} and electromagnetic data~\cite{imitation} (see also~\cite{olivares}).
Moreover, 
macroscopic composites of fermions and bosons, dubbed fermion-boson stars, have also been proposed~\cite{HENRIQUES1990511,jetzer1990,valdez2013dynamical,brito2015accretion,brito2016interaction,valdez2020fermion,Roque:2021}. Such mixed configurations could form from the condensation of some primordial gas containing both types of particles or through episodes of accretion. The dynamical formation of fermion-boson stars through accretion along with their nonlinear stability properties has recently been studied by~\cite{valdez2013dynamical,valdez2020fermion,DiGiovanni:2020frc,DiGiovanni:2021vlu}. In most studies the neutron star is modeled with a polytropic EoS, the only exception being~\cite{kain2021fermion} who employed a realistic EoS. Mergers of fermion-boson stars have also been studied by~\cite{Bezares:2019}.

This paper is organized as follows: Section~\ref{framework} briefly describes the theoretical framework to build equilibrium models of fermion-boson stars. (Further details are reported in~\cite{DiGiovanni:2020frc}.) Section~\ref{sec:results} contains our main results. Finally in Section~\ref{discussion} we discuss our findings and outline possible extensions of this work. 

\section{Framework}
\label{framework}

In our setup the scalar field is assumed to be only minimally coupled to Einstein's gravity. Therefore, fermions and bosons only interact gravitationally, with the total stress-energy tensor being the sum of both contributions, $T_{\mu\nu}= T_{\mu\nu}^{\rm{NS}} + T_{\mu\nu}^{\phi}$,
where (using units with $c=G=\hbar=1$)
\begin{align}
T_{\mu\nu}^{\rm{NS}}&= [\rho(1+\epsilon) + P] u_{\mu}u_{\nu} + P g_{\mu\nu}, \\
T_{\mu\nu}^{\phi}&= - \frac{1}{2}g_{\mu\nu}\partial_{\alpha}\bar{\phi}\partial^{\alpha}\phi - \frac{1}{2} \mu^2\bar{\phi}\phi \nonumber \\
&- \frac{1}{4} \lambda(\bar{\phi}\phi)^2
+ \frac{1}{2}(\partial_{\mu}\bar{\phi}\partial_{\nu}\phi+\partial_{\mu}\phi\partial_{\nu}\bar{\phi}).
\end{align}
The fermionic part involves the fluid pressure $P$, rest-mass density $\rho$, internal energy $\epsilon$, and 4-velocity $u^{\mu}$, with $g_{\mu\nu}$ denoting the space-time metric. The bosonic matter is described by the complex scalar field $\phi$ (with $\bar{\phi}$ being the complex conjugate) and by the particle mass $\mu$ and self-interaction parameter $\lambda$. 

The equations of motion are obtained from the conservation laws of the stress-energy tensor and of the baryonic particles for the fermionic part
\begin{eqnarray} 
\nabla_{\mu}T^{\mu\nu}_{\rm{NS}} = 0, \label{conservation_laws1} \\
\nabla_{\mu}(\rho u^{\mu}) = 0, \label{conservation_laws2} 
\end{eqnarray}
 and from the Klein-Gordon equation for the complex scalar field
\begin{equation}
\nabla_{\mu}\nabla^{\mu} \phi= \mu^2 \phi + \lambda |\phi|^2 \phi\,, \label{Klein-Gordon}
\end{equation}
together with the Einstein equations, $G_{\mu\nu}=8\pi T_{\mu\nu}$, for the spacetime dynamics. Mixed-star models are built using a static and spherically symmetric metric in Schwarzschild coordinates,  
\begin{eqnarray} \label{Schwarzschild_metric}
ds^2 = -\alpha(r)^2 dt^2 + a(r)^2 dr^2 + r^2 ( d\theta^2 + \sin{\theta}^2 d\varphi^2),
\end{eqnarray}
written in terms of two geometrical functions $a(r)$ and $\alpha(r)$. A harmonic time dependence ansatz for the scalar field is assumed, $\phi(r,t) = \phi(r) e^{i\omega t}$, where $\omega$ is its eigenfrequency. Furthermore we consider a static perfect fluid $u^{\mu}= (-1/\alpha,0,0,0)$. In order to construct equilibrium configurations we solve the following set of ordinary differential equations(ODEs), which are obtained from Einstein's equations:
\begin{align} \label{s1}
\frac{da}{dr} & = \frac{a}{2}\left(\frac{1-a^{2}}{r} +4\pi r \biggl[\biggl(\frac{\omega^{2}}{\alpha^{2}}+\mu^{2}+\frac{\lambda}{2}\phi^{2}\biggl) a^{2}\phi^{2}\right.\nonumber \\
		& \left. +\Psi^{2}+2a^{2}\rho(1+\epsilon)\biggl]\right.\biggl),
\end{align}
\begin{align} \label{s2}
\frac{d\alpha}{dr} & = \frac{\alpha}{2}\left(\frac{a^{2}-1}{r} +4\pi r \biggl[\biggl(\frac{\omega^{2}}{\alpha^{2}}-\mu^{2}-\frac{\lambda}{2}\phi^{2}\biggl) a^{2}\phi^{2}\right.\nonumber \\
		& \left.+\Psi^{2}+2a^{2}P\biggl] \right.\biggl),
\end{align}
\begin{align} \label{s3}
\frac{d\phi}{dr} = \Psi ,
\end{align}
\begin{align} \label{s4}
\frac{d\Psi}{dr} & = -\left(1+a^{2}-4\pi r^{2}a^{2} (\mu^{2}\phi^{2}+\frac{\lambda}{2}\phi^{4}\right.\nonumber \\
	&+\rho(1+\epsilon)-P)\biggl)\frac{\Psi}{r}-\left(\frac{\omega^{2}}{\alpha^{2}}-\mu^{2}-\lambda\phi^{2}\right)a^{2}\phi^{2},
\end{align}
\begin{align} \label{s5}
\frac{dP}{dr} = -[\rho(1+\epsilon)+P]\frac{\alpha^{\prime}}{\alpha},
\end{align}
where the prime indicates the derivative with respect to $r$. The system of equations is closed by the EoS for the nucleonic matter. Previous work on fermion-boson stars \cite{valdez2013dynamical,DiGiovanni:2020frc,valdez2020fermion,DiGiovanni:2021vlu} assumed a simple polytropic EoS to build equilibrium models and a $\Gamma$-law EoS for numerical evolutions
to take into account possible shock-heating (thermal) effects.
In this work we improve the microphysical treatment of the fermionic part of the models  and construct new equilibrium solutions described with realistic, tabulated EoS (see Section~\ref{sec:results}). Despite our models are spherically symmetric we can nevertheless apply them to the X-ray milisecond pulsars J0030+0451 and J0740+6620 since the degree of deformation rotation might induce in these objects is negligible \cite{Miller:2019,Riley:2019}.

The set of ODEs~\eqref{s1}-\eqref{s5} is an eigenvalue problem for the frequency of the scalar field $\omega$ which depends on two parameters, namely the central value of the rest-mass density, $\rho_c$, and of the scalar field, $\phi_c$. As in~\cite{DiGiovanni:2020frc} to obtain the value of the frequency for each solution we employ a two-parameter shooting method to search for the physical solution that fulfills the requirement of vanishing $\phi$ at the outer boundary. Once $\omega$ is obtained, we use a 4th-order Runge-Kutta integrator to solve the ODEs and reconstruct the radial profiles of all variables. 

In order to construct physical initial data we must impose appropriate boundary conditions for the geometric quantities and for both the scalar field and the perfect fluid. We require that the metric functions are regular at the origin. We employ Schwarzschild outer boundary conditions, together with a vanishing scalar field. Explicitly, the boundary conditions read
\begin{eqnarray}
&a(0) = 1, \hspace{0.3cm} & \phi(0) = \phi_{c}, \nonumber\\
&\alpha(0) = 1, \hspace{0.3cm} &  \lim_{r\rightarrow\infty}\alpha(r)=\lim_{r\rightarrow\infty}\frac{1}{a(r)},\nonumber\\
& \Psi(0)=0, \hspace{0.3cm} & \lim_{r\rightarrow\infty}\phi(r)=0, \nonumber\\
&\rho(0) = \rho_{c},  & \hspace{0.3cm}  P(0)=K\rho_{c}^{\Gamma}.
\end{eqnarray}

The total gravitational mass of the solutions can be defined as
\begin{equation}\label{mass}
M_{\rm T}=\lim_{r\longrightarrow\infty}\frac{r}{2}\left(1-\frac{1}{a^2}\right),
\end{equation}
which coincides with the Arnowitt-Desser-Misner (ADM) mass at infinity. We define the radius of the fermionic part as the radial coordinate at which the fluid pressure vanishes, $R_{\rm f} = r(P=0)$, which for the Schwarzschild metric coincides with the circumferential radius.  As the bosonic component of our mixed stars does not have a hard surface, the radius of this contribution, $R_b$, is evaluated, as customary, as the radius of the sphere containing $99\%$ of bosonic particles. The particle numbers for both bosons and fermions are computed as in~\cite{DiGiovanni:2020frc}.

\begin{figure}[t!]
\begin{minipage}{1\linewidth}
\includegraphics[width=1.\textwidth]{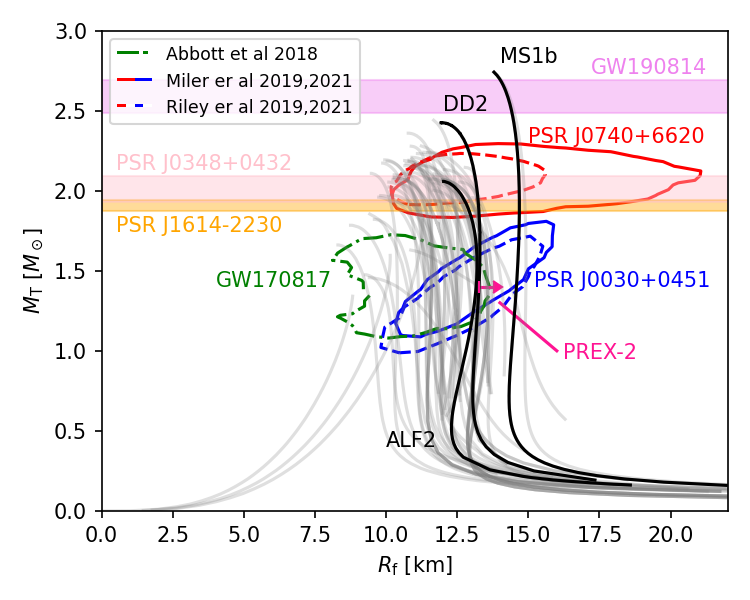}
\caption{Gravitational mass vs circumferential fermionic radius for different realistic EoS including the observational constraints ($95\%$ confidence levels) from LIGO-Virgo, NICER/XMM-Newton, and mass measurements of two high mass pulsars. We also indicate the PREX-2 $1\sigma$ lower limit on the radius for a $1.4~M_\odot$ neutron star. Grey curves correspond to all cold EoS compiled by \cite{compose} and \cite{Ozel:2016}. We highlight in black the three EoS used for the calculations in this work.}
\label{fig:alleos}
\end{minipage}
\end{figure}

\section{Results}
\label{sec:results}

Figure~\ref{fig:alleos} displays the mass-radius relations for a large sample of realistic EoS (grey curves) corresponding to all cold EoS described in \cite{compose} and \cite{Ozel:2016}. Those take into account generic  nuclear effects while some of them also include hyperons, pion and kaon condensates, and quarks. We compare those results with the observational constraints placed by NICER on PSRJ0030+0451~\cite{Miller:2019,Riley:2019}, the NICER/XMM-Newton combined analysis of PSRJ0740+6620~\cite{Miller:2021,Riley:2021}, the constraints set by gravitational-wave event GW170817~\cite{GW170817} (EOS insensitive relations), the mass measurement of two neutron stars with masses close to $2~M_\odot$, PSR J0348+0432~\cite{Antoniadis:2013} and PSR J1614-2230~\cite{PSRJ1614-2230}, and the lower mass component in the binary merger GW190814~\cite{GW190814:2020} as a possible neutron star with mass $\ge 2.5\,M_\odot$. All constraints are given as $95\%$ ($2\sigma$) confidence intervals. Those have been computed using the publicly available posteriors provided by the different groups. Additionally Fig.~\ref{fig:alleos} also shows the $1\sigma$ lower limit for the radius of a $1.4~M_\odot$ derived from the PREX-2 measurements of the neutron skin thickness~\cite{Reed:2021} (see, however, the related discussion in~\cite{Essick:2021,biswas:2021b}).

For our analysis we select the three EoS highlighted in black in  Fig.~\ref{fig:alleos}, namely ALF2, which is a hybrid EoS with mixed APR nuclear matter and colour-flavor-locked quark matter~\cite{Alford_2005}, MS1b which is a relativistic mean field theory EoS~\cite{Alford_2005}, and DD2~\cite{Hempel:2010}, which is a finite temperature hadronic EoS which we evaluate at zero temperature and beta equilibrium. The three EoS fulfill the constraints from the recent NICER and XMM-Newton results, the observations of the two high-mass pulsars, as well as the PREX-2 constraints. Of the three,  only MS1b would be compatible with the low-mass component of GW190814 being a neutron star. On the other hand, only ALF2 and DD2 are compatible with the results of GW170817, albeit only marginally. This selection of EoS illustrates the current tension that exists between different observational and experimental constraints of the mass and radius of neutron stars. Although it is still possible to find EoS that fit all constraints within the $2\sigma$ confidence level (except for GW190814), if these constraints were to tighten in future observations maintaining similar median values, it would pose a serious problem to the modelling of matter at high densities. We explore next the possibility of alleviating some of this tension by considering stars with a bosonic component additional to the fermionic component.

\begin{figure}
\begin{minipage}{1\linewidth}
\includegraphics[width=1.\textwidth]{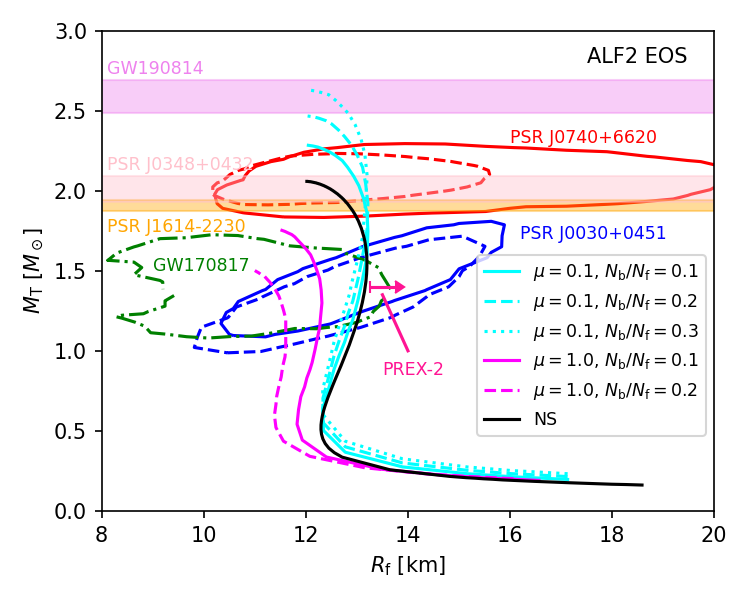}
\includegraphics[width=1.\textwidth]{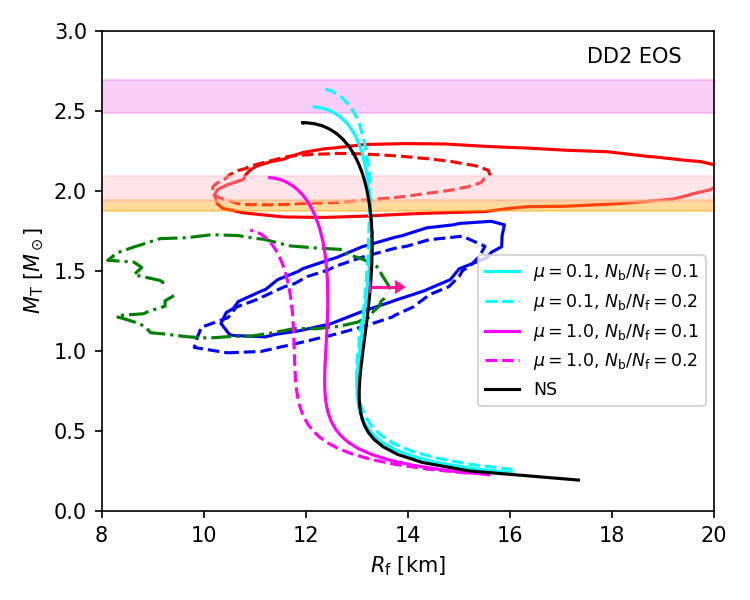}
\includegraphics[width=1.\textwidth]{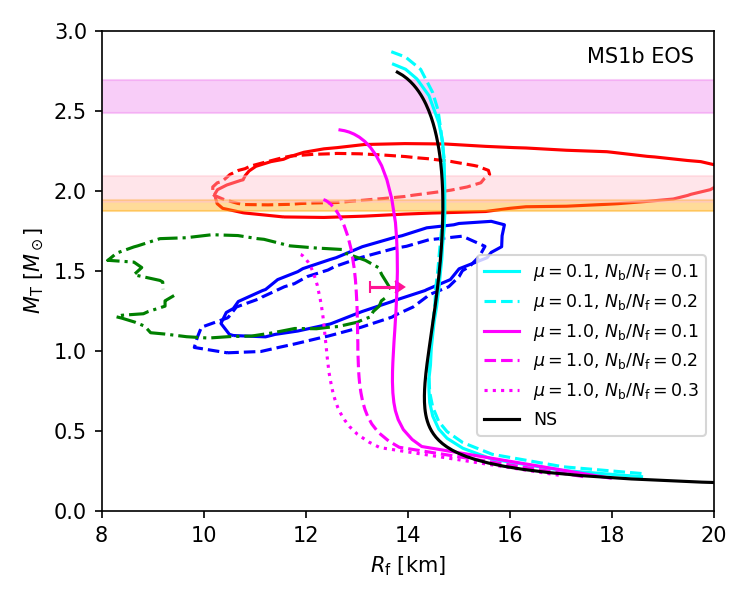}
\caption{Total gravitational mass vs circumferential fermionic radius for equilibrium models of neutron stars (black lines) and boson-fermion stars (magenta and cyan lines) for different parameters of the boson-to-fermion ratio $N_{\rm b}/N_{\rm f}$ and  particle mass $\mu$. The observational constraints plotted are the same as in Fig.~\ref{fig:alleos} and follow the same colour code. Each  panel corresponds to one of the three fermionic EoS described in the text. }
\label{fig:mixed}
\end{minipage}
\end{figure}

With this aim we build sequences of equilibrium configurations of both, fermion stars described by those three EoS, and of mixed stars with different values of the ratio of the number of bosons to fermions, $N_{\rm b}/N_{\rm f}$, and particle mass $\mu$.  Models are computed for $N_{b}/N_{f}=\lbrace0.1,0.2,0.3\rbrace$ and $\mu=\lbrace0.1,1.0\rbrace$ in our units, which correspond to $\mu=\lbrace1.34\times 10^{-11},1.34\times 10^{-10}\rbrace$~eV. For all models the self-interaction parameter $\lambda$ is set to zero (mini-boson stars) and the fermionic matter always dominates over the bosonic matter, the latter being a small fraction of the total mass. The results are depicted in Figure~\ref{fig:mixed}. 

\begin{figure}
\begin{minipage}{1\linewidth}
\includegraphics[width=1.\textwidth]{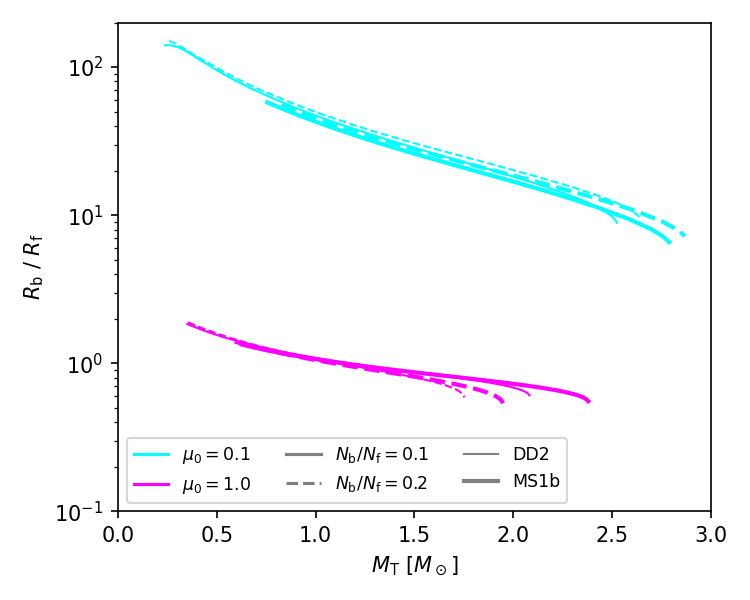}
\includegraphics[width=1.\textwidth]{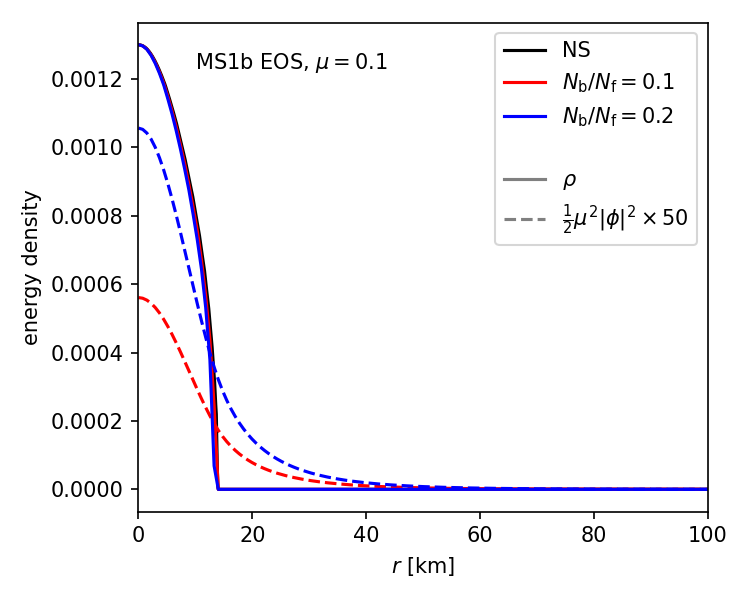}
\includegraphics[width=1.\textwidth]{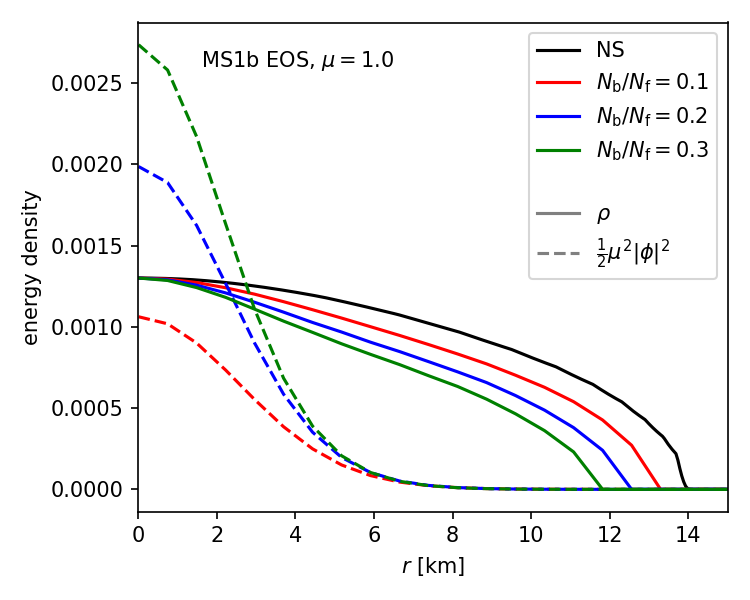}
\caption{Top panel: ratio of the bosonic and fermionic radius as a function of the total mass, for a subset of the models considered in this work. Models not displayed follow a very similar trend. Middle panel:  radial profile of the rest-mass density $\rho$ (solid lines) of an illustrative neutron star model described  by the MS1b EoS and of $\frac{1}{2}\mu|\phi|^2$ (dashed lines) for $\mu=0.1$ and for different values of $N_b/N_f$. Bottom panel: same as middle panel but for $\mu=1.0$.}
\label{fig:radius}
\end{minipage}
\end{figure}

For $\mu=0.1$ (and similarly for smaller values of $\mu$) the size of the bosonic component is larger than the fermionic radius ($R_{\rm b}/R_{\rm f} \sim 10-100$; see cyan curves in the top panel of Fig.~\ref{fig:radius}). In this case the contribution to the gravitational field of the bosonic component, is relatively flat in the region where fermions are present and therefore the impact in the equilibrium configuration of the fermionic component is small.  This results in stars with similar fermionic radii. This is visible in the middle panel of Fig.~\ref{fig:radius} where we show the radial profiles of $\rho$ and $\frac{1}{2}\mu^2|\phi|^2$ for a neutron star and for mixed stars described by the MS1b EoS with the same central value $\rho_c$. However, the additional energy provided by the bosonic component increases the total mass of the system. As a result, in these models (cyan lines in Fig.~\ref{fig:mixed}) the mass of the system increases as the ratio $N_{\rm b}/N_{\rm f}$ increases while keeping the radius almost constant.

On the other hand, for $\mu=1$ (and similarly for larger values of $\mu$) the bosonic component is located in a region similar or smaller than the region occupied by the fermionic component $R_{\rm b}/R_{\rm f} \sim 1$ (see magenta curves in the top panel of Fig~\ref{fig:radius}). In those cases, the bosonic component modifies the gravitational field in the neighborhood of the fermionic component to significantly modify the structure of the star by making it more compact. In those cases (magenta lines in Fig.~\ref{fig:mixed}) the fermionic radius decreases with increasing values of the ratio $N_{\rm b}/N_{\rm f}$ (see bottom panel of Fig.~\ref{fig:radius}). Additionally, due to this increase in compactness, the maximum mass supported by these models decreases.

\section{Discussion}
\label{discussion}

The additional degrees of freedom provided by the presence of a bosonic component may relieve some of the tension observed in the data in several ways. Leaving aside the question of the existence of ultra-light bosonic fields in nature, the main uncertainty of our model is the astrophysical scenario in which fermion-boson stars could form. A number of theoretical works have tried to address this issue (see e.g.~\cite{Brito:2016} and references therein) in particular in the context of ultra-light bosonic fields as a model for dark matter. In order to broadly assess the impact of bosonic fields we explore here two situations that can be regarded as the two limiting cases in the range of possible models.  

{\it 1.~All stars have a constant bosonic-to-fermionic ratio:} the first limiting scenario is the case in which the bosonic field is captured during the formation of the star leading to an approximately universal $N_{\rm b}/N_{\rm f}$ ratio for all fermion-boson stars. In this case, an EoS with relatively low maximum mass for the purely fermionic component, not fulfilling the GW190814 constraint, may produce more massive objects by adding a bosonic component with small values of $\mu$ and solve the issue. Examples are DD2 and ALF2. In these two cases, supplementing a $10-20\%$ amount of bosonic component rises the maximum mass above $2.5~M_\odot$, while preserving the good agreement in radius at lower radii. Note that as a general feature of all EoS (see gray lines in Fig.~\ref{fig:alleos}) the star radius decreases when the maximum mass decreases, meaning that is difficult to have at the same time high maximum masses and small radii. The bosonic contribution is a way of precisely correcting this feature.
Additionally, this procedure can also be used to increase the maximum mass even if in the purely fermionic case this mass is below $2~M_\odot$, which might be a solution to the so-called hyperon problem~\cite{Bedaque:2015}.

{\it 2.~Bosonic-to-fermionic ratio changes over time:} in the second limiting scenario the bosonic matter is assumed to accrete onto the fermionic star after the latter has formed. In this case the ratio $N_{\rm b}/N_{\rm f}$ would increase over time, being higher for older objects. The set of neutron stars considered in this work can be classified in two categories according to their age. Electromagnetically observed pulsars have typical ages smaller than $10$~Gyr; the characteristic age of PSR J0030+0451 is estimated to be $8$~Gyr \cite{Lommen:2000},  PSR J0740+6620 is in the range $5-8.5$~Gyr \cite{Beronya:2019}, PSR J0348+0432 is $2.6$~Gyr \cite{Antoniadis:2013} and PSR J1614-2230 is $5.2$~Gyr \cite{PSRJ1614-2230}. 
On the other hand typical ages of neutron stars found in mergers of compact binaries, such as those in GW170817 and GW190814, may be significantly larger. The merger time for galactic binary neutron star is expected to be in the range $\sim 0.1-1000$~Gyr~\cite{Lorimer:2008}  which is consistent with the estimated merger time of the observed double neutron star systems in the Milky Way~\cite{Tauris:2017}. These estimates should be valid for the two gravitational-wave sources we consider since the metallicity conditions of the host galaxies is likely to be similar to our galaxy, given the low redshift of the sources. For the specific case of GW170817 it has been estimated that the age of the binary must be higher than 1 Gyr~\cite{Levan:2017}. 

Therefore, it is plausible for the second class of objects to have accreted a significantly larger amount of bosonic field and thus have a larger ratio of $N_{\rm b}/N_{\rm f}$ than the first class. In this scenario, neutron star radii could be relatively large for young objects with very small amount of bosonic component, fullfilling the constrains set by PSR J0740+6620 and PSR J0348+0432. And at the same time, potentially older objects such as those in GW170817, would have a significant bosonic component and thus smaller radii (see magenta lines in Fig.~\ref{fig:mixed}). In this situation the bosonic field would need to have a particle mass of at least $\mu=1$. On the other hand, the constraint set by GW190814 would be difficult to fulfill if the secondary was a neutron star, because in this scenario all stars should have much smaller maximum masses, but it could still be explained considering that the secondary is a low-mass black hole. 

Finally, we have to address some of the caveats of our analysis. The observational constraints for the mass and radius considered here assume as a model that the observed object is a neutron star and obtain the posterior distributions according to this model. Therefore, if we change the model by adding a bosonic component, the observational constraints may in principle change as well. For electromagnetic observations of X-ray pulsars, the mass measurement (through Shapiro delay or orbital parameters measurements) relies almost exclusively on the effect of the total gravitational mass, regardless of its composition. The electromagnetic measurement of the radius, on the other hand, determines the size of the observable star, i.e.~the fermionic component alone. The distribution of the bosonic field should affect weakly the analysis of NICER and XMM-Newton because the main effect would be to modify the light bending close to the star (see e.g.~\cite{Riley:2019}). For $\mu=1$ or larger,  most of the bosonic field would be confined inside the fermionic radius. Therefore, the metric outside the observable surface would correspond to that of an object with the total mass of the star and the analysis of NICER/XMM-Newton would be perfectly valid. On the other hand, for $\mu=0.1$ or smaller, most of the bosonic field would be outside the observable surface, and the metric would differ with respect to the one corresponding to the total mass of the system (it would probably be closer to the space-time generated by the fermionic component alone). In that case the analysis of NICER/XMM-Newton would require corrections. 

We also recall that in all of our models the self-interaction parameter of the bosonic field has been set to zero. It would be interesting to study the effect of self-interactions ($\lambda\neq0$) on mixed fermion-boson stars with a realistic EoS since a  self-interaction potential allows to increase the maximum mass without changing the particle mass $\mu$. We leave this analysis as future work.

Regarding gravitational-wave observations, the measured component gravitational masses would probably be well estimated since the structure of the compact objects appears only at 5PN order in the waveform models for binaries~\cite{Blanchet:2014}. However, the estimation of the radius, as done in GW170817, may require modifications. This is actually an indirect estimation as the actual  parameter measured is the quadrupole tidal deformability. From this, assuming that the object is a neutron star, it is possible to put constrains on the radius~\cite{Read:2013}. Therefore, to do a proper analysis one should have to either make the relevant corrections to estimate the fermion-boson star (fermionic) radius from the observational constrains on the tidal deformability or to compute the tidal deformability of our mixed stars (in particular the quadrupole Love number) to compare directly with observations. Either of the two analysis is out of the scope of this paper. However, even if we do not  perform this analysis we expect that the trends found in our work should at least be qualitatively correct since there is a correlation between the tidal deformability and the radius.

It is also worth noticing that in scalar-tensor theories of gravity, in which the scalar field is not minimally coupled to gravity, neutron star models present significant deviations from general relativity through spontaneous scalarization, leading to neutron stars with significantly larger masses and radii~\cite{Damour:1992,Harada:1998,Hajime:2004,Degollado:2020}. In this regard, a suitable choice of the scalar field parameters and coupling constants of scalar-tensor theories could effectively reproduce the same mass-radius relations we have discussed in this paper for mixed fermion-boson stars in general relativity. Such potential degeneracy would make difficult to distinguish between the two cases and, thus, between the underlying theory of gravity.

On a similar note, while our model  resembles those of~\cite{Das:2021,Lee:2021} we have applied it to explain a larger set of observational and experimental data than those authors, who exclusively focused on explaining the secondary component of GW190814 as a potential dark-matter-admixed neutron star. Our findings for GW190814 agree with those of~\cite{Das:2021,Lee:2021} which provides an independent consistency check. Since in our model the bosonic component plays the role of dark matter, it is not surprising that any similar dark-matter model would likely fit the data, irrespective of the type of matter considered.

To summarize, we conclude that the addition of a bosonic component to a neutron star leads to mixed configurations with mass-radius relations that are compatible with recent multi-messenger observations of compact stars, both in the electromagnetic channel (PSR J0030+0451 and PSR J0740+6620) and in the gravitational-wave channel (GW170817 and GW190814), as well as with the latest PREX-2 experimental results. The possibility of enlarging the parameter space of neutron stars with different contributions from the bosonic component offers thus a theoretically-motivated approach to reconcile the tension in the data collected by NICER/XMM-Newton and the LIGO-Virgo-KAGRA Collaboration.

\begin{acknowledgements}
We thank Reed Essick, Nick Stergioulas, and Juan Carlos Degollado for useful comments. Work supported by the Spanish Agencia Estatal de Investigaci\'on (grant PGC2018-095984-B-I00), by the Generalitat Valenciana (PROMETEO/2019/071 and GRISOLIAP/2019/029), by the European Union’s Horizon 2020 RISE programme (H2020-MSCA-RISE-2017 Grant No.~FunFiCO-777740), by DGAPA-UNAM (grants No.~IN110218, IA103616, IN105920), and by the Funda\c c\~ao para a Ci\^encia e a Tecnologia (FCT) (projects PTDC/FIS-OUT/28407/2017, CERN/FIS-PAR/0027/2019, and PTDC/FIS-AST/3041/2020). This work has further been supported by FCT through Project~No.~UIDB/00099/2020. P. C.-D. acknowledges support from the Spanish Ramon y Cajal programme (RYC-2015-19074). We acknowledge networking support by the COST Action GWverse CA16104.
Computations have been performed at the Lluis Vives cluster of the  Universitat de Val\`encia and the Argus and Blafis clusters of the Universidade de Aveiro. We also  gratefully acknowledge the Italian Instituto Nazionale di
Fisica Nucleare (INFN), the French Centre National de la Recherche Scientifique (CNRS)
and the Netherlands Organization for Scientific Research, for the construction and operation
of the Virgo detector and the creation and support of the EGO consortium.
\end{acknowledgements}

\bibliography{num-rel2}

\end{document}